\documentclass[aps,prl,twocolumn,superscriptaddress,showpacs,a4paper,amsmath,amssymb,byrevtex]{revtex4}

\usepackage[latin1]{inputenc}
\usepackage{graphicx}
\usepackage{bm}
\usepackage{times}

\begin{document}

\title{Non-universal behavior of the parity effect in monovalent atomic wires}

\author{P{\'e}ter Major}
\affiliation{Department of Solid State Physics, E{\"o}tv{\"o}s
University, 
H-1117 Budapest, P\'azm\'any P{\'e}ter s{\'e}t\'any 1/A,
Hungary }

\author{V.~M. Garc\'{\i}a-Su\'arez}
\affiliation{Department of Physics, Lancaster University, Lancaster,
LA1 4YB, UK}

\author{S.~Sirichantaropass}
\affiliation{Department of Physics, Lancaster University, Lancaster,
LA1 4YB, UK}

\author{J{\'o}zsef Cserti}
\affiliation{Department of Physics of Complex Systems, E{\"o}tv{\"o}s
University, H-1117 Budapest, P\'azm\'any P{\'e}ter s{\'e}t\'any 1/A,
Hungary}

\author{C.~J.~Lambert}
\affiliation{Department of Physics, Lancaster University, Lancaster,
LA1 4YB, UK}

\author{J. Ferrer}
\affiliation{Departamento de F\'{\i}sica, Facultad de Ciencias, 
Universidad de Oviedo, 33007 Oviedo, Spain}

\author{G{\'e}za Tichy}
\affiliation{Department of Solid State Physics, E{\"o}tv{\"o}s
University, 
H-1117 Budapest, P\'azm\'any P{\'e}ter s{\'e}t\'any 1/A,
Hungary }


\begin{abstract}

We propose a mixed analytical-ab-initio method for the accurate
calculation of the conductance in monovalent atomic wires.
The method relies on the most general formula for ballistic transport 
through a monovalent wire, whose parameters can be determined from 
first-principles calculations. Our central result is the demonstration
of the highly non-universal behavior of the conductance, which depends 
on the fine details of the contacts to the leads. We are therefore able
to reconcile a large number of the apparently contradictory results that 
have recently appeared in the literature.   

\end{abstract}

\pacs{73.23.-b,73.63.Rt,73.40.Cg}

\maketitle

In the last two decades transport properties of atomic contacts 
have been the subject of intensive research 
(for overviews see Ruitenbeek~\cite{Schon1} and 
Agra{\"\i}t, Yeyati and Ruitenbeek~\cite{Ruitenbeek_overview:cikk}). 
First experimental evidences of the formation of golden atomic chains
have been reported by Yanson et al. 
and Ohnishi et al.~\cite{Yanson-Ohnishi:cikk}. 
Experiments on chains of Au, Pt and Ir atoms~\cite{Smit:cikk} exhibit 
electrical conductance oscillations as a function of the wire length
and similar oscillations as a function of bias voltage 
and electrode separation~\cite{Ludoph:cikk,Halbritter-1}.
Rodrigues et al.~\cite{Rodrigues:cikk} investigated 
the energetically preferred orientation of 
the crystal planes of the wire by the application of high-resolution 
transmission electron microscopy.  
Their results show a strong correlation between the 
atomic arrangement and the conductance. 

The above experiments were stimulated by early theoretical 
predictions of conductance quantization~\cite{Jaime:cikk} and conductance
oscillations~\cite{Pernas:cikk,Lang:cikk}. 
The latter issue generated a sequence of theoretical
papers using a variety of 
techniques~\cite{Lang_2:cikk,Emberly:cikk,Lang_3:cikk,Kobayashi:cikk,Sim:cikk,Gutierrez:cikk,Lee_Kim:cikk,Zeng:cikk,Havu:cikk,Tsukamoto_Hirose:cikk,Thygesen:cikk,Hirose_Kobayashi:cikk,Lee:cikk}. 
Density-functional theory predicted that the conductance of Na atom
chains is close to the conductance quantum $2e^2/h$ for odd numbers
of atoms, and smaller than this for even numbers of 
atoms~\cite{Lang:cikk,Sim:cikk,Gutierrez:cikk,Tsukamoto_Hirose:cikk}.  
In the literature this is called the even-odd effect. 
A similar effect was found for other 
monovalent alkali-metal atoms such as Cs, but an 
opposite behavior, with a conductance bigger for even numbers of
atoms than for odd numbers of atoms, was predicted for 
noble-metals (Cu, Ag and Au)~\cite{Lee:cikk}.  
The even-odd oscillation of the conductance for atomic wires of Na 
has also been analyzed using a pseudoatom-jellium
model~\cite{Havu:cikk}, where it was found that the sign of the effect 
is sensitive to the lead cone angle. 
Applying the first-principle recursion-transfer-matrix method Hirose at
al.~\cite{Hirose_Kobayashi:cikk} showed that 
the bonding nature of the atoms at the contact 
plays a crucial role in determining transport properties.
Lee and Kim\cite{Lee_Kim:cikk} and Zeng and Claro\cite{Zeng:cikk},
studied the effects of symmetries and found that the
conductance of atomic chains with mirror symmetry and an odd number of atoms
is always equal to the conductance quantum $2e^2/h$. 

In the literature various heuristic models have been proposed 
to interpret physically the results mentioned above:
the standing wave model proposed by Emberly et al.~\cite{Emberly:cikk},
a simple barrier model suggested by Lee et al.~\cite{Lee:cikk},  
a simplified one-dimensional free-electron model of Smit et 
al.~\cite{Smit:cikk}, the Friedel sum rule and charge
neutrality of Sim, Lee et al.~\cite{Sim:cikk,Lee_Sim:cikk} and, more
recently, a resonant transport model used by Thygesen and Jacobsen to
explain the four-atom period oscillation of the conductance for Al
wires~\cite{Thygesen:cikk}.

To reconcile the different behaviors observed in the literature and
elucidate whether the parity behavior is universal or
depends on the fine details of the system, we
present in this letter a theoretical scheme which
treats electronic structure and transport to
the same degree of rigor and interprets results in terms of a simple,
but general framework.  We use the non-equilibrium
Green's function approach, combined with a rather general, model-free
formula, where a derivation requires only two assumptions:   
i) the existence of a {\em translational invariant} region 
within the atomic chain, 
ii) the presence of only {\em one conducting channel} in the wire, even though 
many conductance channels are allowed in the leads. 
We then demonstrate the validity of this formula 
by comparing with the conductance calculated from ab-initio calculations 
for different choices of the contacts.  

\begin{figure}[hbt]
\includegraphics[scale=0.65]{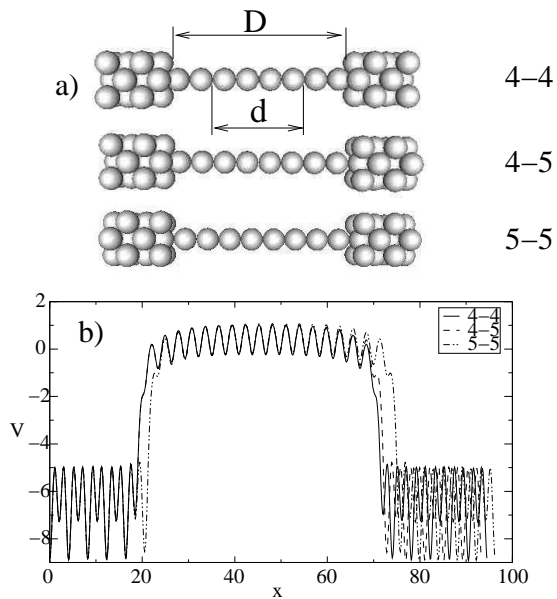}
\caption{(a) Three different couplings of the chain to the leads were 
studied: the chain coupled to both leads by 4 atoms (case 4-4), 
the chain coupled to the left and right lead by 4 and 5 atoms,
respectively (case $4-5$), and 
the chain coupled to both leads by 5 atoms (case $5-5$).  
(b) The variation of the potential $V$ (in units of Rydbergs) 
along the Au chain with 18 number of atoms (in units of \AA ) 
in the case of the three types of couplings shown in (a).
\label{fig:geo_pot} }
\end{figure}

Figure~\ref{fig:geo_pot} (a) shows atomic chains of physical length $D$,
containing a translationally-invariant region of length $d=(n+1)a$, 
where $n$ is the number of atoms in that region and $a$ is the lattice
constant of the chain. 
In general, $D>d$ because atoms near the contacts may experience a
different electrostatic potential from those in the region of length $d$. 
To verify this we performed atomistic calculations of the electronic
structure of gold and sodium chains using the density functional code
SIESTA~\cite{Koh65,Sol02,siesta}. 
Gold and sodium are prototypes of noble and alkali metals,
respectively.
 The external leads used in the simulations are bcc for sodium and fcc for
gold, both grown along the (001) direction, and can be terminated 
with a surface cross-section containing either 4 or 5 atoms.
This allows us to perform three separate calculations corresponding
to three different choices of termination of the left-right leads,
which we denote 4-4, 5-5, and 4-5. 
We also used leads with bigger cross sections and leads grown along
(111) for gold and verified that the overall predictions are 
essentially the same.
Figure~\ref{fig:geo_pot} (b)
shows the electrostatic potential $V$ as a function of the distance
$x$ along the chain of gold. One can see a rapid variation of
the potential near the contact and an almost perfect periodic
dependence around the middle of the chain in a relatively long
translationally invariant segment of length $d$.

The system comprising the translational invariant segment of the atomic 
chain connected to atoms on the left and right can be characterized by 
two scattering matrices that couple the left and the right regions to
the segment $d$. 
Using the unitary relations for the scattering matrices and 
the Landauer-B\"uttiker formula~\cite{Landauer-Buttiker-Datta:ref}
we find that the conductance (in units of $2e^2/h$) is 
\begin{equation}
g =  \frac{\left(1-r^{\prime 2}_L \right) 
\left(1-r^{2}_R \right)}
{1+ r_{L}^{\prime 2} r_{R}^2
- 2 r_{L}^{\prime} r_{R} 
\cos \left(2kd + \Phi_{LR} \right)}, 
\label{G_scattering:eq}
\end{equation}
where  $ \Phi_{LR} = \Phi_{L} + \Phi_{R}$ and 
$\Phi_L$($\Phi_R$), and $ r^\prime_L $($r_R$) 
are the phase shifts and magnitudes of reflection 
amplitudes at left (right) ends of the region $d$ and $k$ is
the wavenumber of the transmitting electron.
This model-free equation is the most general form of the conductance 
through a monovalent ballistic atomic chain. 
Note that in (\ref{G_scattering:eq}) $r^{\prime}_L$, $r_R$ 
and $ \Phi_{LR}$ are scalars depending on the details of the coupling
of the region $d$ to the leads and the Fermi energy,  
and are independent of the length $d$ of the chain.
Thus, the conductance is a periodic function of the length $d$.  
As we shall demonstrate, the parameters $r^{\prime}_L$, $r_R$ and 
$\Phi_{LR}$ can be determined from ab-initio calculations. 

We immediately notice two differences between 
Eq.~(\ref{G_scattering:eq}) and the corresponding result  
obtained from a model based on the simplified one-dimensional free-electron 
picture~\cite{Smit:cikk} and the barrier model~\cite{Lee:cikk}. 
First, $d$ is the length of the translational invariant region of the 
chain, which is not necessarily equal to $D$.
Second,  Eq.~(\ref{G_scattering:eq}) contains a phase, which is 
missing from the analysis of Refs.~\onlinecite{Smit:cikk,Lee:cikk} 
and even for a model in which $D=d$, this phase is in general, non-zero.
Consequently, the even-odd behavior of the conductance strongly depends 
on the details of the coupling of the chain to the leads
(through the phase shift $\Phi_{LR}$) and this 
is the reason for the different predictions in the 
literature~\cite{Smit:cikk,Lee:cikk,Havu:cikk,Emberly:cikk,Zeng:cikk,Lang:cikk}. 
It is obvious from (\ref{G_scattering:eq}) that the conductance $g$ 
becomes unity when the two leads are identical ($r^{\prime}_L = r_R$),
and the cosine term is 1. 
For the simple tight-binding approximation used 
in, e.g.,~\cite{Todorov:cikkek} the dispersion relation is 
$\varepsilon(k) = \varepsilon_{0} + 2  \gamma  \cos(k  a)$, 
where $\varepsilon_{0}$ and $\gamma$ are the on-site energy 
and the hopping energy. 
For a chain of monovalent atoms the Fermi energy is 
$\varepsilon_0$ and the Fermi wave number is $ k =\pi/(2a)$.
Hence, for $\Phi_{LR}=0$ and for an odd number of atoms $n$, i.e., $d=(n+1)a$, 
the conductance is $g=1$, and is smaller for even $n$, 
in agreement with~\cite{Zeng:cikk,Lee_Kim:cikk} 
and the model in~\cite{Lee:cikk,Smit:cikk}. 
However, this prediction is in contrast to 
the results of~\cite{Emberly:cikk} for gold wires and~\cite{Lee:cikk} 
for noble-metals, implying that in those cases the phase 
$\Phi_{LR} \ne 0$. 

To obtain the conductance from an ab-initio calculation 
we used our newly developed code SMEAGOL~\cite{Roc04,Natmat}, 
which calculates the density matrix and the transmission coefficients
of a two probe device from the Hamiltonian provided by SIESTA using the
non-equilibrium Green's function formalism (NEGF)~\cite{Kel65}.
In this calculation the main input is the atomic number and
atomic positions of the constituents, whereas
Eq.~(\ref{G_scattering:eq}) is obtained from a model-free scattering
approach.
The ab-initio calculation yields $g$ as a function of the energy
$\varepsilon$ and does not compute $r^{\prime}_L$, $r_R$ 
and $ \Phi_{LR}$ directly. 
To obtain the latter, we compute the
ab-initio conductance $g(\varepsilon)$ for a range of $\varepsilon$
and perform a least squares fit to Eq.~(\ref{G_scattering:eq}),
treating  $r^{\prime}_L$, $r_R$, $ \Phi_{LR}$ and 
$k(\varepsilon)$ as fitting parameters.

\begin{figure}[hbt]
\includegraphics[scale=0.58]{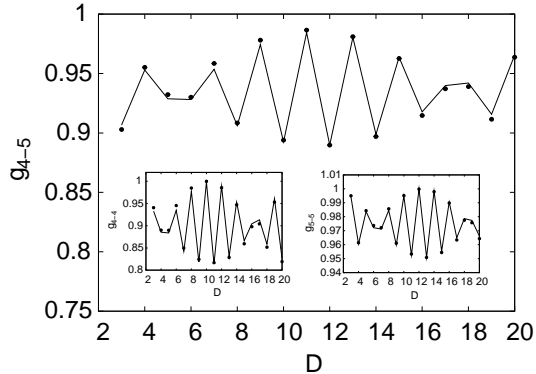}
\caption{The main panel shows the conductance $g$ 
(in units of $ {2e^{2}}/{h}$) obtained from the ab-initio calculations (dots)
of Na chains between different leads (4-5) 
as a function of length of the chain  (in units of the number of atoms)
for $ \varepsilon =E_{F}+ 0.099 $ eV, and the conductance (solid line) 
obtained by substituting $ r'_{L}=r_{44} $, $ r_{R}=r_{55} $ 
and $ \Phi_{LR}=(\Phi_{44}+\Phi_{55})/2 $ in Eq.~(\ref{G_scattering:eq}).
In the left (right) insets the results of the ab-initio calculations
(dots) and the fits of $g$ (solid line) using
Eq.~(\ref{G_scattering:eq})  
for the 4-4 (5-5) cases are shown yielding the parameters 
$ r_{44}$, $ \Phi_{44} $, and $ r_{55} $, $ \Phi_{55} $, respectively. 
\label{fig:check} }
\end{figure}

At first sight, it is not obvious that the simplicity of 
Eq.~(\ref{G_scattering:eq}) is sufficient to capture the complexity of
an ab-initio calculation. 
To show the validity of this formula we now determine the 
parameters in Eq.~(\ref{G_scattering:eq}) from atomistic calculations 
and show that the phase $\Phi_{LR}$ of the asymmetrical system 
satisfies a simple addition rule in the following sense.
For identical leads with a particular choice of contact 
denoted by $\alpha$ the parameters $r^{\prime}_L = r_R$ 
and $\Phi_{LR}$ obtained from ab-initio calculations will be
denoted by $r_{\alpha \alpha} $ and $\Phi_{\alpha \alpha} $, respectively.
For a symmetric system with another choice of contact to the leads,
say $\beta$, these parameters will be denoted 
by  $r_{\beta \beta} $ and $\Phi_{\beta \beta} $. 
Equation~(\ref{G_scattering:eq}) requires that 
the conductance of an asymmetrical system with the left-lead contacted 
as $\alpha$ and the right-lead as $\beta$, is obtained by substituting 
$r^{\prime}_L = r_{\alpha \alpha} $, $r_R = r_{\beta \beta} $ 
and $\Phi_{LR}=(\Phi_{\alpha \alpha} + \Phi_{\beta \beta})/2$ 
into formula (\ref{G_scattering:eq}). 
Figure~\ref{fig:check} shows the predicted conductance of the
asymmetrical lead configuration using the parameters obtained from 
the two symmetrical cases for Na chains (similar results are obtained
for gold).  
As can be seen in the main panel of the figure the agreement is
excellent. 

\begin{figure}[hbt]
\includegraphics[scale=0.62]{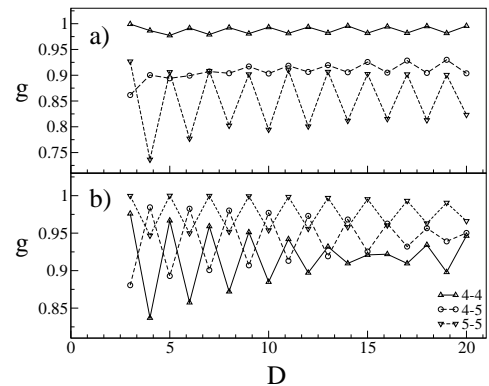}
\caption{The conductance (in units of $2e^2/h$) obtained from atomistic
simulations for Au (a) and Na (b) atomic chains 
at energy $\varepsilon = E_{F} $ 
as a function of length of the chain $D$ (in units of the number of
atoms) for the three different choice of contacts shown 
in Fig.~\ref{fig:geo_pot}a. 
\label{fig:cond} }
\end{figure}

\begin{figure}[hbt]
\includegraphics[scale=0.52]{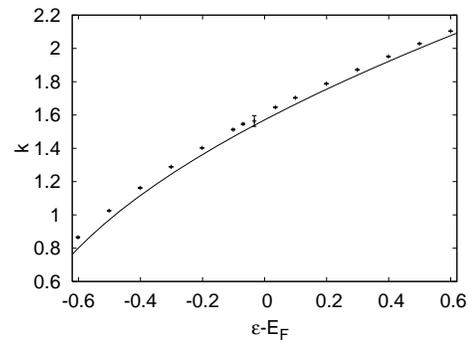}
\caption{The energy-wave number relation obtained with help of formula
(\ref{G_scattering:eq}) ($+$ sign and error bars) and dispersion relation calculated using SIESTA (solid line) for an infinite Na atomic chain. The energy
and the wavenumber are in units of eV and the inverse of the lattice
constant, respectively.
\label{fig:disp} }
\end{figure}

Figure~\ref{fig:cond} shows the ab-initio results at the Fermi energy
for the conductances of Na and Au chains. 
One can see that  the conductances strongly depend
on the material and the
type of coupling, demonstrating thus the non-universal behavior of
the even-odd effect in these systems.
In each case the parity effect is obvious, which means that wavenumber 
$k$ in Eq.~(\ref{G_scattering:eq}) at the Fermi energy 
is near to $ {\pi}/{(2a)}$.
We also compared the predicted $k(\varepsilon)$ with the band
structure obtained from a separate ab-initio calculation on an
infinitely-periodic atomic chain.
Our results of $k(\varepsilon)$ from  fitting and the ab-initio
band structure calculation for an infinite long  
Na atomic chain are shown in Fig.~\ref{fig:disp}.  
The good agreement is clearly visible. 

\begin{figure}[hbt]
\includegraphics[scale=0.44]{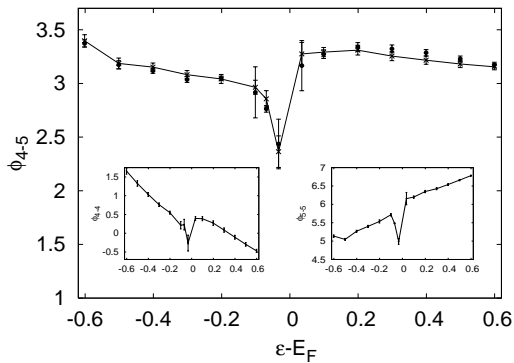}
\caption{The insets show the phase  $\Phi_{44}$ and $\Phi_{55}$ 
(in units of radians) for Na chains. 
The main figure shows a comparison between 
$\Phi_{45}$  ($ \times $ sign and line with error bars) and the average 
$(\Phi_{44}+\Phi_{55})/2$ (filled circle with error bars). 
The energy scale is in units of eV.
\label{fig:addfi} }
\end{figure}

The validity of the addition rule for the
phases can be checked directly, as is demonstrated in
Fig.~\ref{fig:addfi} for sodium wires. One can see from the figure that
the agreement is again very good (only close to the Fermi energy the
errors of the numerical fit are larger). 
Similar agreement is also obtained for gold chains, not shown here.
We find that there are significant differences in the phase shifts 
$\Phi_{LR}$ for different lead-wire configurations. 
The results shown in Fig.~\ref{fig:addfi} demonstrate that 
the phase of the even-odd behavior is sensitive to the lead-chain 
coupling and can be opposite to that which is normally expected. 

Equation~(\ref{G_scattering:eq}) also explains why the
ensemble-averaged conductance of Au wires~\cite{Smit:cikk} 
oscillates with small amplitude, which in most cases
is less than about 0.05 in unit of $2e^2/h$.
Neglecting the sample dependence of the reflection amplitudes 
$r^{\prime}_L$, $r_R$ in formula (\ref{G_scattering:eq})
and assuming a uniform distribution for the phase shift $ \Phi_{LR}$
over the interval $[0,2\pi]$,  
one can calculate the ensemble-averaged conductance by integrating 
formula (\ref{G_scattering:eq}) with respect to $ \Phi_{LR}$, 
to yield an average conductance, which is independent of
the length $d$ of the chain. 
Since the experimental result for the average conductance shows 
a small, but finite amplitude of oscillation, the distribution of the
phase $ \Phi_{LR}$ is presumably not exactly uniform, 
or correlated to sample-dependent reflection amplitudes. 
This suggests the importance of a sample-dependent 
statistical analysis of the calculated or measured conductance for 
a better understanding of the conductance oscillation.
We think that formula (\ref{G_scattering:eq}) through the 
functional form of the conductance can be useful starting point  
for such an analysis.

In summary, a simple, but model-free formula
for the transport through monovalent atomic chains was presented 
and compared with the results obtained from ab-initio calculations 
for Au and Na based on the non-equilibrium Green's function approach. 
Our results clearly demonstrate that i) Eq.~(\ref{G_scattering:eq})
describes accurately the conductance through atomic wires with a variety
of contacts provided there is a translationally invariant segment
in the chain with one conducting channel and ii) the non-universal
behavior of the parity effect, which helped us to understand and
reconcile the different results observed in the literature.

We believe that the conductance formula ~(\ref{G_scattering:eq})
derived from the scattering approach, unifies and captures the essential
feature of the earlier 
models~\cite{Emberly:cikk,Lee:cikk,Smit:cikk}.
Moreover, as it is known, the resonant assisted transport 
such as developed 
in Refs.~\onlinecite{Sim:cikk,Lee_Sim:cikk,Thygesen:cikk}, in general, 
can be interpreted by the scattering matrix approach  
(see e.g.\ Ref.~\onlinecite{reso-Colin:cikk} and references therein).

\acknowledgments 

We would like to thank Gy.~Mih\'aly and A.~Halbritter for useful
discussions.
This work is supported by E.~C. Contract No.~MRTN-CT-2003-504574, EPSRC,  
the Hungarian-British TeT, and the Hungarian Science Foundation OTKA 
Grant Nos.~TO34832 and T32417. VMGS thanks the European Union for a Marie
Curie grant.


\begin{thebibliography}{99}
\bibitem{siesta} SIESTA uses pseudopotentials to get rid of the core
electrons and linear combinations of atomic orbitals to span the valence.
We used a double$-\zeta$ polarized basis set for both Na and Au and the
LDA functional to calculate exchange and correlation.
\end{thebibliography}
\end{document}